%% file: pn2b.tex
\documentclass[10pt,a4paper]{llncs}
\usepackage[english]{babel}
\usepackage[T1]{fontenc}
\usepackage{ae}
\usepackage{multicol}

\usepackage{pstricks}
\usepackage{fancybox}
\usepackage[dvips]{graphics}
\usepackage{zedter}
\usepackage{zed-csp}
\usepackage{diagrams}
\RequirePackage{times}
\usepackage{boxedminipage}
\usepackage{multicol}

\title{Semantic Embedding of Petri Nets into Event-B}
\medskip 

\author{J. Christian Attiogb{\'e} \\[0.7ex]  
}
\institute{LINA  - FRE CNRS 2729\\
2, rue de la Houssini{\`e}re, B.P.92208, F-44322 Nantes Cedex 3, France\\
\texttt{Christian.Attiogbe@univ-nantes.fr}\\
}

\begin{document}
\maketitle
\begin{abstract}
We present an embedding of Petri nets into B abstract systems. 
The embedding is achieved by translating both the static structure (modelling aspect) and the evolution semantics of Petri nets.
The static structure of a Petri-net is captured  within a B abstract system through a graph structure. This abstract system is then included in another abstract system which captures the evolution semantics of Petri-nets. The evolution semantics results in some B events depending on the chosen policies: basic nets or high level Petri nets. 
The current embedding  enables one to use conjointly Petri nets and Event-B in the same system development, but at different steps and for various analysis.
\end{abstract}
\section{Introduction}
Reliable system development requires the use of concepts, languages, tools and methods which are provided by formal approaches. Several mono-paradigm methods already exist.
However, real size systems often overwhelm the scope covered by mono-paradigm specification techniques and  their complexity requires an adequate integration of appropriate techniques and methods for both the development and the formal analysis. 
Current research efforts focus on  the combination of various approaches and their specific tools in order to strengthen their impact on industrial system treatment. 
Therefore, there are some requirements to make formal methods more practical and efficient in their usage: 
\textit{i)} they should be linked with \textit{engineering practices} and techniques; 
\textit{ii)} their {\it mechanization} by providing powerful and {\it operational development tools}. These points are still some challenges for the formal method community and therefore they motivate our work. 

The integration of different formal methods may be motivated by different kind of combinations: 
the complementarity of methods so as to cover the facets of the application at hand,
the need of specific techniques such as composition and refinement, 
or specific reasoning techniques such as theorem proving and model checking,
or pragmatic considerations such as graphical formalisms and the interoperability of tool supports.

In the current work we study the integration of Petri nets and B in order to use conjointly both approaches in the same development. 
The motivation is to benefit from the complementarity of both approaches. 
Petri nets formalism may be used as a graphical front-end of a B development project. 
The B framework may follow to complement formal analysis of the system modelled using Petri nets.\\
On the one hand, Petri nets formalism are widely used  \cite{murata89,PNET12_1998,reisig98,KristensenPN2004} by engineers and also in academic or research projects. 
Petri nets also have graphical facilities, simulation and  liveness property verification facilities via powerful model checking techniques.
On the other hand, B is a model-based approach which permits correct development with refinement from abstract specifications to executable codes; it is based on theorem proving technique and it offers (mainly) safety properties verification facilities. 

The contribution of this article resides in 
\textit{i)} the definition of a (B) generic structure to capture Petri nets models and semantics; 
\textit{ii)} the means to systematically embed Petri nets structure and their evolution rules into Event-B. 
This leads to the development of a bridge between Petri nets and B.

The article is organised as follows: 
in  Section 2 we introduce the Petri nets formalism and the B Systems approach. 
Section 3 is devoted to the stepwise embedding of Petri nets into B: basic nets are first considered and then generalized to high level nets.
Section 4 gives some issues related to analysis and in Section 5 we give some concluding remarks. 

\section{Petri Nets and B Systems}
\subsection{An Overview of Petri Nets (P-nets)}
Formally, a P-net is a 4-tuple $(P,~ T,~ Pre,~ Post)$ where :
\begin{itemize}
\item $P$ is a  finite set of places , (with $|P|~=~m$, the cardinal of $P$);
\item  $T$ is a finite set of transitions,  (with $|T|~=~n$, the cardinal of $T$);
\item $P$ and $T$ are disjoint sets  ($P \cap T = \{ \}$);
\item  $Pre~:~P~\times~T~\rightarrow~\nat$ is an input function, $Pre(p,t)$ denotes the number of arcs from the place $p$ to the transition $t$;
\item  $Post~:~P~\times~T~\rightarrow~\nat$ is an output function,  $Post(p,t)$ denotes the number of arcs from the transition $t$ to the place $p$.
\end{itemize}

Practically, a P-net is a bipartite directed graph whose arcs connect nodes from two distinct sets; the set of places and the set of transitions. Petri nets are equipped with a graphical formalism where the places are connected to the transitions using the directed arcs.\\

\noindent
\textit{Graph associated to a P-net.}
The graph  associated to a net  $N$ is described by:
\begin{itemize}
\item $\Gamma_p$ the transitions reachable from each place:\\
 $\forall p ~\in~P~.~\Gamma_p(p)~=~\{ t~\in~T~|~Pre(p,~t)~>~0\}$ 
\item $\Gamma_t$ the places reachable from each transition:\\
 $\forall t ~\in~T~.~\Gamma_t(t)~=~\{ p~\in~P~|~Post(p,~t)~>~0\}$ 
\item $W_{in}$ the weight of each input arc: $\forall~p~\in~P,~\forall~t~\in~T~.~W_{in}(p,~t)~=~Pre(p,t)$ and 
\item $W_{out}$ the weight of each output arc: $\forall~p~\in~P,~\forall~t~\in~T~.~W_{out}(p,~t)~=~Post(p,t)$
\end{itemize}
The graph associated to a P-net is the abstract representation which is used to manipulate the net. 
The places connected to a transition with an arc from each place to the transition are the \textit{input places} of the transition.
The places connected to a transition with an arc from the transition to each place are the \textit{output places} of the transition.\\

\noindent
\textit{P-net Marking.} 
A marked net $M_N=(N,~ \mu)$ is made of a net $N$ and a  mapping $\mu ~:~P~\rightarrow~\nat$.\\
$\mu(p)$ is the number of tokens within $p$; it is called the \textit{marking} of the place $p$. 
The initial marking $M_0$ of a net is the n-tuple made of the initial marking of all the places $p_i$ of the net: $M_0 =(\mu(p_1), \cdots,\mu(p_m))$ where $m$ is the number of places.\\

\noindent
\textit{Behaviour of a P-net.}
A P-net evolves by firing some \textit{enabled} transitions. 
A transition is \textit{enabled} if all its input places contain at least so many tokens as is the weight of the arcs  from the place to the transition. 
An enabled transition may be fired and enable all the actions in the output places of the transition.  There is a nondeterministic choice between the enabled transitions.
Firing a transition modifies the markings of both input and output places. 
This may enable or disable other transitions. All enabled transitions may be fired.
Therefore the  evolution of the net describes a \textit{marking net} which can be infinite.
When a transition is  fired, one token is removed from every input place of the transition and 
one token is added to every output place of the transitions. This is generalized by removing (resp. adding) the number of tokens corresponding to the weight of the arcs from the input place to the transition (resp. to the weight of the arcs from the transition to the output place). 

\subsection{An Overview of B Abstract Systems}
\label{section:overview_EB}
\input{ovv_eventB.tex}
\section{Embedding Petri Nets into Event-B}
\subsection{Embedding techniques}
Embedding techniques are introduced in \cite{BoultonEtAl92} and provide a methodology to reuse existing logical frameworks for formal analysis.
Embedding techniques are intensively used for method integration and mechanization of notations \cite{BowenGordon95,GravellPrattenZ_PVS_HOL,Munoz&Rushby:FM99}. 
There are two main embedding  techniques: {\it shallow embedding} and {\it semantic embedding} (also called {\it deep embedding}). 
The first technique deals with the translation of specifications (objects of a formalism) to semantically equivalent objects in the target formalism. Nevertheless, the mapping from the language constructs to their semantic representations is part of  the metalanguage (support of the source language).
In the case of semantic embedding, the complete semantics of a source formalism is translated into the target formalism: both syntax and semantics of the source language are formalized inside the target language logic. That means, the mapping from language constructs to their semantic representations is part of the target language logic. 
Consequently, using semantic embedding, we do not need only the (semantic preserving) syntactic translation of the constructs but also the semantics to be translated into the target logic. The choice of one of the techniques depends on the envisaged goal.
\subsection{Embedding the Structure of Petri Nets within B}
Embedding the  structure of a P-net into B (Fig. \ref{figure:structRdP}) consists in describing the graph associated to the P-net. 
The 4-tuple which describes a net $N$ is encoded with the set of places ($places$), 
the set of transitions ($transitions$), and the two relations between places and transitions ($placesBefore, placesAfter$).  
Additionally we have the  marking functions for the places: $mu$.
We also consider the weights of the arcs; they are  natural number greater or equal to the unit.
The input arc weights are described by the function $weightBefore$.
The output arc weights are described by the function $weightAfter$.
Therefore some invariant properties may be added.
This results in an event-less B abstract system (Fig. \ref{figure:structRdP}) which captures only the graph structure of a marked net $(N, mu)$.\\ 
It remains to deal with the behavioural semantics of the Petri net.
This is based on the marking of the net and the transitions. 
\begin{figure}[htp]
\begin{center}
\noindent
\begin{boxedminipage}{6cm}
\begin{tabbing}
\hspace{0.4cm}\=\hspace{0.4cm}\=\hspace{0.4cm}\=\hspace{0.4cm}\kill
\textsc{system} \textit{PetriNet}\\
\textsc{sets} \\
 \>PLACE;  ~~ TRANSITION\\
\textsc{variables } \\
\>$places, transitions, placesBefore, placesAfter, weightBefore, weightAfter, mu$\\
\textsc{invariant} \\
\>$places$ $\subseteq$ PLACE \\
$\land$\> $transitions$ $\subseteq$ TRANSITION \\
$\land$\> $placesBefore \in transitions \rel places$ ~~~~/* $placesBefore^{-1} = \Gamma_p$~~  */\\
$\land$\> $placesAfter  \in transitions \rel places$ ~~~~~~/* $placesAfter = \Gamma_t$  ~~~~~~~~*/\\
$\land$\> $placesBefore = \dom(weightBefore)$ \\
$\land$\> $placesAfter = \dom(weightAfter)$ \\
$\land$\> $weightBefore \in transitions \cross places \pfun \nat$\\
$\land$\> $\dom (weightBefore) = placesBefore$\\
$\land$\> $weightAfter \in transitions \cross places \pfun \nat$\\
$\land$\> $\dom (weightAfter) = placesAfter$\\
$\land$\> $mu : places \fun \nat$
\end{tabbing}
\end{boxedminipage}
\end{center}
\caption{A Partial B system encoding a P-net}
\label{figure:structRdP}
\end{figure}
\subsection{Embedding Petri Nets Evolution Semantics into B}
\label{section:embedevol}
A P-net evolves by firing the enabled transitions. From a given marking, firing one of the enabled transitions, leads to a new marking of the net and so on. This is embedded in event-B by an abstract system whose events correspond to the transition firing. 

A P-net transition may be  formalized (at first approximation) as a B event (see Fig. \ref{figure:dynamicsRdP}) with a guard which expresses that all the input places of the transition have the required number of tokens and  a body (a generalized substitution) which expresses the update of input places (by removing the necessary tokens) and the update of output places (by adding the appropriate number of tokens). B events are instantaneous and their effect can cause the occurrence of other events. This copes well with the semantics of P-net: the firing of a transition $t_i$ is instantaneous and thus can lead to the firing of other transitions which have the output places of $t_i$ among their input places.
\vspace{-0.4cm}
\subsubsection{Basic Petri net}
Here \textit{basic} Petri net means that  actions (data+operations) are not attached  to the places nor to the transitions.
The arc weight may be greater or equal to the unit.
The guard for firing a transition is that all its input places have the required number of tokens:  $\forall p \in placesBefore(t).~ \mu(p) \ge weightBefore(t, p)$.
The effect of firing a transition is the update, via the $\mu$ function, of the input and output places according to the input and output arcs:\\
\centerline{ $\forall p \in placesBefore(t). \mu(p) := \mu(p) - weightBefore(t, p)$}
\centerline{ $\forall p \in placesAfter(t). \mu(p) := \mu(p) + weightAfter(t, p)$}
Therefrom,  the firing of a transition $t_i$ is translated with a single B event \textbf{\texttt{event\_tr}} (Fig. \ref{figure:dynamicsRdP}) which works for every transition $t_i$ in a nondeterministic way. The variables $mupbef$ and $mupaft$ model with B generalized substitutions the update of the $\mu$ function as described above. The notation $s \dres r$ expresses the restriction of the domain of the relation $r$ to the elements in the set $s$. 
Likewise $<\!\!\!\!+$ denotes the overriding of a relation by another one.\\
To simplify the reading, we take some freedom with the notation of the abstract systems given in the remainder of the article.   
\begin{figure}[htp]
\begin{center}
\noindent
\begin{boxedminipage}{5.7cm}
{\small
\begin{tabbing}
\hspace{0.0cm}\=\hspace{0.1cm}\=\hspace{0.4cm}\=\hspace{0.4cm}\kill
\> \textbf{\texttt{event\_tr}} $\defs$ ~~~~~~~~~~~~\textsf{/* for any transition t$_i$ */} \\
\>\>    \textsc{any}  $t_i$, $pbef$, $paft$, $pb, vv, pa, uu, mupbef, mupaft, \cdots$ \textsc{where}\\
\>\>\>       $pbef = placesBefore[\{t_i\}]$ 
$\land~paft = placesAfter[\{t_i\}]$\\
\>\>$\land  mupbef \in places \fun NAT \land mupbef \subseteq mu \land dom(mupbef) = pbef$\\ 
\>\>$\land pb \in pbef \land vv \in NAT$\\
\>\>$\land vv+weightBefore(ti,pb) < MAXINT$\\
\>\>$\land ( ((pb, vv) \in mupbef) \implies (pb \in pbef \land (pb, vv+weightBefore(ti,pb)) \in (pbef \dres  mu)))$\\
\>\>$\land  mupaft \in places \fun NAT \land mupaft \subseteq mu \land dom(mupaft) = paft$\\ 
\>\>$\land pa \in paft \land uu \in NAT$\\
\>\>$\land uu-weightBefore(ti,pa) > 0$\\
\>\>$\land ( ((pa, uu) \in mupaft) \implies (pa \in paft \land (pa, uu-weightAfter(ti,pa)) \in (paft \dres  mu)))$\\
\>\> $\land$ \> $\cdots$ \\
\>\>    \textsc{then} \\
\>\>\>   /* \textit{update the  places before and after  the transition} $t\_i$ */\\
\>\> \> $ mu := mu <\!\!\!\!+ (mupbef \cup mupaf)$ \\
\>\>$\|$\>   $\cdots$ \\
\>\>    \textsc{end}
\end{tabbing}
}
\end{boxedminipage}
\end{center}
\caption{A shape of a B event capturing the evolution of a basic P-net}
\label{figure:dynamicsRdP}
\end{figure}

We captured the behavioral semantics of basic P-nets with a B abstract system with a \textit{single event} representing the transitions of the net. This abstract system simulates the evolution of the P-net.  
Using a single event for all transitions instead of one event per transition simplifies the generalisation and the reasoning on the embedding; indeed only the structure of a parameter P-net needs to be translated for each new project.
\vspace{-0.4cm}
\subsubsection{Generic Structure of the Embedding}
We show in Figure \ref{figure:structurEmbed} the B generic structure which holds all P-net model; it is the abstract system named $EmbeddedPN$. We separate the encoding of the semantics ($EmbeddedPN$) which works for any P-net and the static structure part ($PetriNet$) which is specific to a problem and should be included for a given problem.
The static part (in the $PetriNet$ abstract system) is increased with some variables: $pl\_actions$ is the set of actions attached to the places. 
The injective (total) function\footnote{It is injective because we need the reverse function.} $pl\_treatment \in places \inj pl\_actions$ records the action located in each place; a specific element $nullaction$ is used for the initialisation and for action-less places.\\  
The system $EmbeddedPN$ has two variables: the relation $trans\_places$  records, for the currently fired transition(s), the output places which are not yet processed; 
the function $guard\_P\_actions$  is used to get the guard of each place action.\\
The single event \texttt{\textbf{event\_tr}} manages the firing of transition and thus the evolution of the considered net. This event is improved and replaced in the following sections by two (or several events according to the considered policy) related events (\texttt{action\_ak, fire\_transition}).
\noindent 
\begin{figure}[htp]
\begin{center}
\noindent
\begin{boxedminipage}{6cm}
\begin{tabbing}
\hspace{0.4cm}\=\hspace{0.4cm}\=\hspace{0.4cm}\=\hspace{0.4cm}\=\hspace{8cm}\kill
\textsc{system} \textit{EmbeddedPN}\\
\textsc{includes} \\
 \>  PetriNet~~~~~~~~~/* any described P-net; this is a parameter */\\
\textsc{variables } \\
\> $guard\_P\_actions, trans\_places$\\
\textsc{invariant} \\
\>$guard\_P\_actions \in pl\_actions \fun BOOL$\\
$\land$\> $trans\_places \in transitions \rel places$ \\
\textsc{initialisation} \\
\>$guard\_P\_actions := ((pl\_actions-\{nullaction\}) \times \{FALSE\})$\\
\>\>\>\>$\cup \{(nullaction, TRUE)\}$\\
$\|$ \> $trans\_places := \{\}$ \\
\textsc{events} \\
\> \texttt{event\_tr} $\defs$~~ $\cdots$~~~~~\textsf{/* for any transition ti */} \\
\textsc{end}
\end{tabbing}
\end{boxedminipage}
\end{center}
\caption{Generic Structure of the Embedding}
\label{figure:structurEmbed}
\end{figure}

Therefrom we extend the embedding to cover more complicated cases: action management. Indeed, according to their types (place/transitions, conditions/event, resources, etc), P-nets may deal with data and actions (or treatments) in different manners. 

\begin{figure}[htb]
\centerline{\resizebox{5.6cm}{4.0cm}{\includegraphics{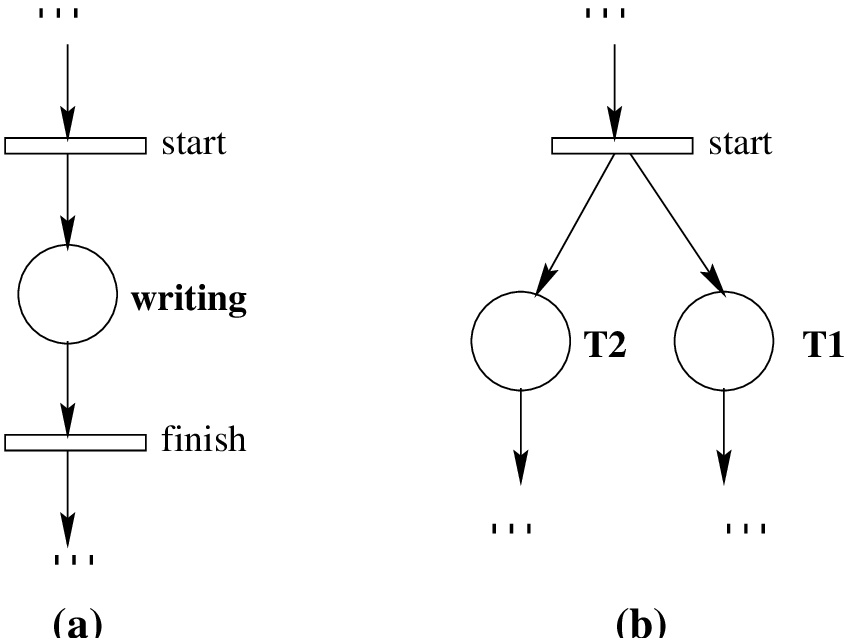}}}
\caption{Places with associated action}
\label{figure:treatmentab}
\end{figure}

In some P-nets the places with tokens may model availability of data; in this case an action may be associated to the transition.
In other models, some places may contain  action which is then guarded by one or several transitions. It is for instance the case in a net modelling a process writing some data in exclusion with other writer processes; a specific place is often used in such a case (see Fig. \ref{figure:treatmentab} (a)).
Therefore there is not a single way to embed the P-nets. We investigated both cases of action attachment: to the places (e.g. \textbf{writing}, see Fig \ref{figure:treatmentab} (a)) and to the transitions (e.g. \textbf{T1, T2}, see Fig. \ref{figure:treatmentab} (b)).  
\subsection{Dealing with Non-Basic Nets}
In the previous section, we considered the evolution of basic P-nets; no specific policies or treatments  are considered.
\vspace{-0.4cm}
\subsubsection{High Level Petri Nets}
High Level Petri Nets (HLPN) were introduced to overcome the problem of the explosion of the number of elements needed for large computer systems. 
HLPN use 
\textit{i)}  structured data to model the tokens, and algebraic expressions to annotate the net elements;
\textit{ii)}  transition modes to describe more elaborated operations/actions.\\
Within HLPN the enabling of a transition depends not only on the availability of the tokens but also on their nature. 
There are several achievements of HLPN \cite{Jensen9296};\\
  Predicate/Transition-Nets \cite{GenrichPrTNets_86} and Colored Petri-Nets \cite{JensenCPN1,appliPN2004} are two forms of HLPN.
In this article, we consider an abstraction of the ideas of HLPN.  Actions (treatments or operations) may be associated to places and transitions of the nets. This corresponds to the idea of structured tokens, typed places and typed transitions, and more generally the execution of some operations associated to the places or to the transitions of a net. Accordingly, we have a generic treatment of the whole.
 
The study is achieved step by step; first we examine the formalisation in the case where actions are associated to places only. Then we study the cases where they are associated to transitions. Finally we consider the general case where actions are associated to both the places and the transitions. 
\vspace{-0.4cm}
\subsubsection*{Petri Nets with Actions Attached to Places}
The actions attached to the places should be achieved when the places are guarded by a transition which is fired.
Thereby \textit{each action in a place of a P-net is translated as  a (guarded) event of the B abstract system}. \\
In practice, actions need some time to be completed.
Therefore firing a transition may be achieved in two steps: 
\textit{i)} enabling the guard of all the  actions attached to the output places of the transition; 
\textit{ii)} launching nondeterministically these \textit{involved actions}. All of them should be performed in any order.

\begin{figure}
\centerline{\resizebox{5.8cm}{4.4cm}{\includegraphics{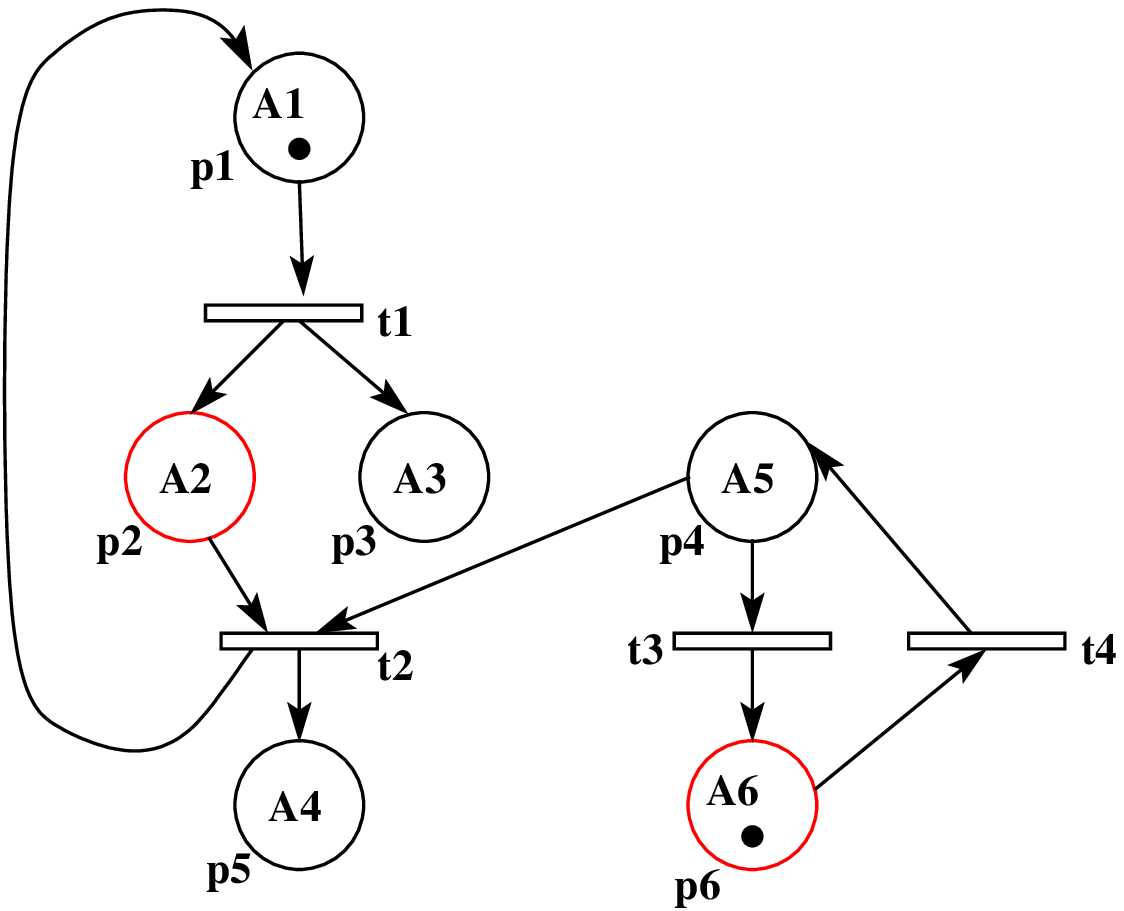}}}
\caption{Interdependent Actions}
\label{figure:treatmentc}
\end{figure}
This raises some questions: what about the duration of the actions and the enabling of other transitions?
Should we wait for the completion of an action before considering another action?
What about the scheduling of the enabled transitions and enabled actions?
Considering these questions  with respect to Figure \ref{figure:treatmentc} gives an idea of the complexity of the scheduling of actions;
the transition t1  enables the actions \{A2, A3\};
 t2  enables the actions \{A4, A1\};
 t3  enables the action \{A6 \};
 t4  enables the action \{A5 \}.\\
These actions are interdependent because the places that contain them are either an input place or an output place of the fired transitions.
There are cycles; for example, firing repeatedly the transitions t3 and t4. \\
To deal with the current situation, we use the previously defined (see Section \ref{section:embedevol}) variables $pl\_treatment$, $pl\_actions$, $guard\_P\_actions$. The firing of a transition $ti$ is handled with two events which correspond to the two steps distinguished above.\\
The first step of the transition firing is captured  with the B event \textbf{fire\_transition\_tr} given in Figure \ref{figure:dynamicsRdP1}.
The output places of a transition $t_i$ are: $paft = placesAfter[\{ti\}]$.
The involved actions  associated to these places are: $involved\_actions = paft \dres pl\_treatment$. 
  
The  guards of the involved actions attached to the output places  of the fired transition are enabled ($\forall~Ai~\in~placesAfter[\{ti\}].~ guard(Ai) := true$).
The  function $guard\_P\_actions$ is updated in order to enable the guards. 
This is done with a Cartesian product: $ran(involved\_actions)\times\{TRUE\}$. 
The marking of the input places is updated.
The fired transition and its output places are recorded in the relation $trans\_places$; this is necessary for the scheduling of involved actions. Indeed \textit{all the actions of the output places should be performed before the actions of the possible transitions they can enable}.
\begin{figure}[htp]
\begin{center}
\noindent
\begin{boxedminipage}{6cm}
\begin{tabbing}
\hspace{0.4cm}\=\hspace{0.4cm}\=\hspace{0.4cm}\=\hspace{0.4cm}\kill
\> \textbf{fire\_transition\_tr} $\defs$ ~~~~~~~~~~~~\textsf{/* for any transition ti */} \\
\>\>    \textsc{any}  $tr_i$, $pbefs, pbef$, $mupbef, pa, vv, \cdots$ \textsc{where}\\
\>\>\>       $pbefs =  placesBefore[\{t_i\}]$ $\land$ $paft = placesAfter[\{t_i\}]$\\

\>\>$\land$ \>$pbef = \{plc | plc \in pbefs \land mu(plc) \ge weightBefore(t_i, plc) \}$ \\
\>\>$\land$ \>$involved\_actions = paft \dres pl\_treatment$\\
\>\>$\land$ \> $\cdots$ \\
\>\>    \textsc{then} \\
\>\>\>   $guard\_P\_actions := ran(involved\_actions)\times\{TRUE\}$ \\
\>\>\>~~~~~ /* enabling the guard of involved actions */\\
\>\>$\|$\> $ mu :=  mu <\!\!\!\!+~ mupbef$\\
\>\>\>~~~~~  /*  update only of the input places of  $t_i$ */ \\
\>\>$\|$\> $trans\_places := trans\_places \cup (\{t_i\} \times paft)$\\
\>\>\>~~~~~  /*  the output places of $t_i$; they will be updated later on */ \\
\>\>    \textsc{end}
\end{tabbing}
\end{boxedminipage}
\end{center}
\caption{Dynamic part of  the generic structure (a)}
\label{figure:dynamicsRdP1}
\end{figure}

Since the B events are atomic we cannot update the marking of output places during the first step; they will eventually enable other transitions which will take place.
Moreover, to cope with practical application of P-nets, one has to consider the "run until completion" of the various actions during their scheduling.

The second step of the firing is captured with the event \textbf{\texttt{action\_Ak}} (see Fig. \ref{figure:dynamicsRdP2}).
One B event is described for each action associated to a place. 
This enables us to handle the high level aspect of the net; indeed the treatments depend on the tokens and on the transitions. 
The guard of each action is maintained until the action is performed.
The actions attached to the output places which are still enabled, are nondeterministically performed; they are recorded in (the range of) $trans\_places$. But, the actions in the places contained in $trans\_places$ can be performed at any time (due to the nondeterminism of event occurrence). 
When an action is completed its guard is disabled and the number of tokens of the place is updated: the function  $trans\_places$ is updated,
the $mu$ function is  updated to set the marking of output places. 

\begin{figure}[htp]
\begin{center}
\noindent
\begin{boxedminipage}{6cm}
\begin{tabbing}
\hspace{0.4cm}\=\hspace{0.4cm}\=\hspace{0.4cm}\=\hspace{0.4cm}\kill
\> \textbf{\texttt{action\_Ak}} $\defs$ ~~~~~~~~~~~~\textsf{/* for an action Ak (attached to a place pp) */}\\
\>\>    \textsc{any} $pp, tr, weiga, \cdots$  \textsc{where} ~~~~\textsf{/* pp is the  place of the action Ak  */}\\
\>\>\>   $pp \in PLACE \land pp = pl\_treatment^{-1}(Ak)$ \\
\>\>$\land$ \>  $guard\_P\_actions(Ak) = TRUE$\\
\>\>$\land$ \> $(tr, pp) \in trans\_places$ \\
 \>\>$\land$ \> $weiga = weightAfter(tr,pp)$\\
\>\>$\land$ \> $\cdots$\\
\>\>    \textsc{then}\\
\>\>\>     $\cdots~~~~$ \textsf{/* place to put an effective action Ak */}\\
\>\>\>     $guard\_P\_actions(Ak) := FALSE$\\
\>\>$\|$\> $mu(pp) := mu(pp) + weiga$\\
\>\>$\|$\> $trans\_places := trans\_places -\{(tr,pp)\}$\\
\>\>    \textsc{end}
\end{tabbing}
\end{boxedminipage}
\end{center}
\caption{Dynamic part of the generic structure  (b)}
\label{figure:dynamicsRdP2}
\end{figure}

However, there are some shortcomings with the current case.
There is a kind of loss of priority between actions: if the effect of one of the currently enabled actions contributes to fire another transition, the actions enabled by this latter transition can be performed before the actions already enabled (this comes fatally from the substitution $trans\_places := trans\_places \cup (\{t_i\}\times paft)$).\\
Another shortcoming is the following: when there are cycles, an enabled guard (of an action) can be overwritten; that is, the enabling condition can be observed again whereas the already enabled action is not yet performed.

We solve these problems in the general case presented later on, by using priorities.
\vspace{-0.4cm}
\subsubsection*{Petri Nets with Actions Attached to Transitions}
In the same way as for the previous case with places, a total function $tr\_treatment \in transitions \fun tr\_actions$ records the action associated to each transition.\\
$tr\_actions$ is the set of actions attached to all the transitions; it is defined in the static structure ($PetriNet$). 
When an enabled transition is fired, its associated action should be performed before the update of the marking of the output places, otherwise another transition may take the priority over the current. \\
Several transitions may share the same input place(s). But, when the latter has the necessary number of tokens to enable the transitions which share the place, only one of the enabled transitions is fired. 
Therefore two steps are necessary to handle the firing of a transition.
In a first step, one of the enabled transitions is nondeterministically selected; the guard of the action associated to this transition is enabled. The marking of all the input places is updated. This is quite similar to the event \textbf{fire\_transition\_tr}.
In a second step, the transition action is performed; its guard is disabled and, the marking of the output places is updated. These places may enable other transitions and so forth.
We get two B events corresponding to the described steps:
\textit{i)} a  firing event which is used to select a transition and to update the input places; this event deals with all the enabled transitions;
\textit{ii)} each transition action has an event with its associated guard which depends on the marks of input places.
\vspace{-0.4cm}
\subsubsection*{Petri Nets with Actions Attached to both Places and Transitions}
In the current case, when a transition is fired, the action linked with it is enabled and the marking of the  output places of the transition is updated; these output places have actions which should be enabled. 
After that, the transition action is performed, it enables the actions linked to the output places. Moreover, the action linked to the places should be performed before enabling the transitions linked to them.
In order to embed this semantics, we use additionally with the preceding variables, the function $enabled\_P\_actions$ for the currently enabled place actions; $enabled\_T\_actions$ for the currently enabled transition actions. We remind that $trans\_places$ records which output places are not yet processed for the currently fired transition.  

The embedding is achieved according to priority rules.
The priority between actions are handled as follows.
A transition is fired if 
\textit{i)} the input places have the required number of tokens, 
\textit{ii)} there is no previous enabled place action not yet performed;
this is checked with $(trans\_places=\{\})$.
Indeed when a transition is fired, its action is enabled and it enables some (output) place actions. They all should be performed before firing another transition. This policy solves the problem of guard overwriting.

Therefrom the event \textbf{fire\_transition\_tr} is modified  as described in Figure \ref{figure:dynamicsRdP3_PT}.
\begin{figure}[htp]
\begin{center}
\noindent
\begin{boxedminipage}{6cm}
\begin{tabbing}
\hspace{0.4cm}\=\hspace{0.4cm}\=\hspace{0.4cm}\=\hspace{0.4cm}\kill
\> \textbf{fire\_transition\_tr} $\defs$ ~~~~~~~~~~~~\textsf{/* for any transition ti */} \\
\>\>    \textsc{any}  $t_i$, $pbefs, pbef$, $mupbef, pa, vv, \cdots$ \textsc{where}\\
\>\>\>       $pbefs = placesBefore[\{t_i\}]$ $\land paft = placesAfter[\{ti\}]$\\
\>\>$\land$ \>$pbef = \{plc | plc \in pbefs \land mu(plc) \ge weightBefore(ti, plc) \}$ \\
\>\>$\land$\> $\textbf{trans\_places = \{\}}$ $\land$ $\textbf{(enabled\_P\_actions $\rres$ \{ TRUE \}) =\{\}}$\\
\>\>$\land$\> $involved\_actions = paft \dres pl\_treatment$\\
\>\>$\land$ \> $\cdots$ \\
\>\>    \textsc{then} \\
\>\>\>   $enabled\_T\_actions(t_i) := TRUE$ \\
\>\>\>~~~~~ /* enabling the  action of the transition */\\
\>\>$\|$\>   $enabled\_P\_actions := ran(involved\_actions) \times \{TRUE\}$\\
\>\>\>~~~~~ /* enabling the guard of the involved place actions */\\
\>\>$\|$\> $ mu := mu <\!\!+ mupbef$ ~~/*  update of the input places of  $t_i$ */\\
\>\>$\|$\> $trans\_places :=  \{t_i\} \times paft$\\
\>\>\>~~~~~  /*  get the output places of $t_i$; they will be updated later on */ \\
\>\>    \textsc{end}
\end{tabbing}
\end{boxedminipage}
\end{center}
\caption{Dynamic part of a Petri Net with  Place and Transition Actions}
\label{figure:dynamicsRdP3_PT}
\end{figure}

The remaining accompanying events (not detailed here) are the following:\\
\textbf{enable\_transition\_action\_guard}; it sets the guard of an enabled transition action to TRUE, then it disables the transition guard.\\
\textbf{enable\_place\_action\_guard}; it sets the guard of an enabled place action to TRUE, updates the $mu$ function and updates $trans\_places$ by removing the treated  place;\\
\textbf{launch\_transition\_action\_aj}; it launches one of the transition action whose guard is $true$ and then it sets the guard to $false$;\\
\textbf{launch\_place\_action\_ak}; this one launches a place action whose guard is enabled, then the guard is disabled.

All these five events (of the abstract system $EmbeddedPN$) simulate an interleaving run of the firing of transition actions and place actions, but priority is employed to avoid bad behaviour of the actions. 
The entire system is checked for consistency using Atelier B; some analysis issues are considered with various case studies.  
\section{Analysis Issues}
\vspace{-0.2cm}
\subsection{Analysis of Petri Nets}
Very often, two classes of properties are studied on P-nets:
one is about the boundedness of the nets. For example the accumulation of tokens in a place is symptomatic of a bad functioning of a model. 
The second class is about the liveness of the nets. By studying the reachability of certain marking, one can detect deadlock freedom for example. 
In all these cases, the marking graph (the set of reachable markings) should be computed. This aspect of the analysis may raise some  problems.The size of the graph may be too large to be analysed in a reasonable time; the graph may also be infinite. When the graph is infinite, a covering graph is used; it enables to check a part of the desired properties.  

Three main class of analysis techniques \cite{murata89,PNET12_1998} for P-nets are: \\
\textit{Reachability analysis}: it is based on state space exploration/reduction techniques using model checking. The main idea is to construct an occurrence graph (a directed graph) which has a node for each reachable system state (a marking) and an edge for each possible state transition. The analysis is then based on such graph.\\
Reachability is like a simulation of the modelled system execution. It allows for a rapid analysis of the system to check for its functionalities. \\
\textit{Structural analysis}:  algebraic analysis are applied here. \\
\textit{Invariant analysis}: it consists to check that some properties associated to the places are satisfied for all reachable states (a net marking) of the modelled system.  

The advantages of the first analysis techniques are: 
a graph is constructed and analysed systematically; 
the constructed graph may be very large; but it exists techniques which makes it possible to work with minimized graphs.
However the main disadvantage is that, such a graph may become very large, even for very small systems, rending the analysis unpractical due to state explosion problem.

One of the aspects on which this work contributes in is the definition of the basis for the combined use of analysis techniques and tools. The available B platforms may be used to analyze the safety properties of systems which are modelled with P-nets. 
\vspace{-0.3cm}
\subsection{An Illustration: Producer-Consumer with Semaphore}
\begin{figure}
\centerline{\resizebox{8.5cm}{7.1cm}{\includegraphics{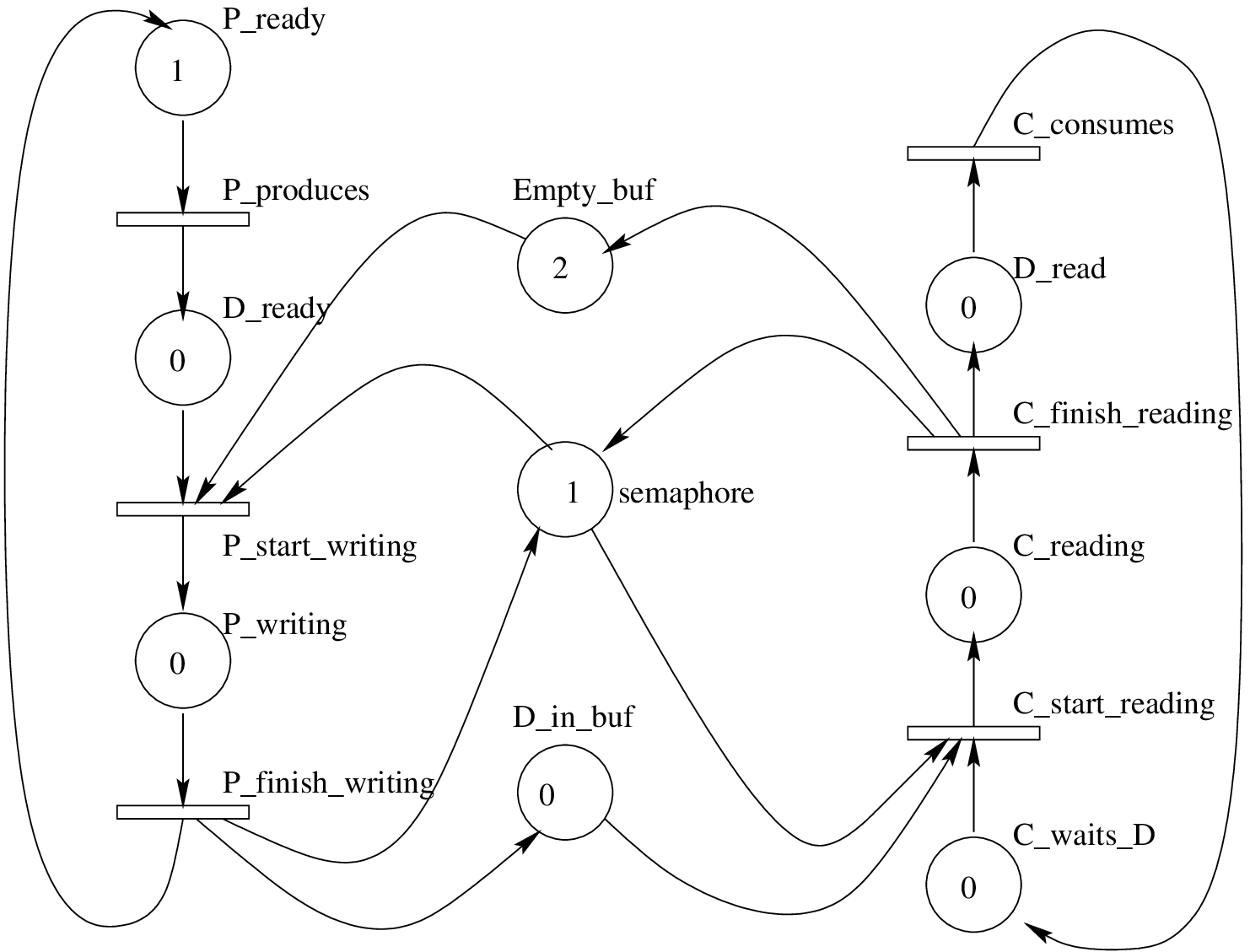}}}
\caption{A producer consumer example}
\label{figure:prodcons}
\end{figure}
We described and checked the producer-consumer system depicted in Figure \ref{figure:prodcons} using our approach. Only the description of the abstract system $PetriNet$ is given; it is included in the system $EmbeddedPN$ which does not change (it gathers the P-nets semantics). 
Additionally to the properties that may be analyzed in a standard Petri-net platform, some safety properties that may be analysed using the B tools are:
\vspace{-0.2cm}
\begin{itemize}
\item Boundedness of some places: the places Empty\_buf and D\_in\_buf (see Fig. \ref{figure:prodcons}) are bounded. This is formalized as the following predicate which is added to the invariant:\\
 $mu(Empty\_buf) \le 2 \land mu(D\_in\_buf) \le 2$
\item There is not a bad usage of the resources (here the buffer):\\
 $mu(Empty\_buf) + mu(D\_in\_buf) = 2$
\item The system is \textit{live}; that means there is always at least one transition which can be fired; this is formalized with:\\
$placesBefore\verb|~|[dom(nmu \rres \{ii | ii \in \nat \land ii > 0\})]  \neq \{\}$
\end{itemize}
This illustrates how we may manage the modelling and analysis work through Petri-nets and B. We achieved some experiments with such properties. In the current article, due to lack of space we do not go into the details of this analysis aspect.
\section{Conclusion and Further Work}
We  presented an embedding of Petri nets formalisms (and modelling) into the B abstract system formalism. The embedding is systematic and it covers basic P-nets as well as high level nets.
The current work fills a gap between the widely practiced P-nets formalism and the emerging proof-based development technologies especially the B method which is based on abstract machines, refinement and theorem proving.
That is a step towards a multi-facet analysis framework for relating discrete system modelling techniques.\\
\textit{Results.}  We provide  a two-level embedding infrastructure made of a generic B abstract system that may be used to describe any Petri-net and, an abstract system that includes (genericity) the first one and whose events capture the semantics of Petri-nets evolution.
Various policies concerning high level P-nets have been considered.
Concretely we may combine the use of P-nets and B method in the same project; for example we may begin the modelling with an existing graphical tool dedicated to the P-nets and then follow with the B method for some related aspects.
This work is generally related to works on embedding techniques but it is specifically related to the work by Sekerinski and Zurob \cite{SekerinskiZurob02} on Statecharts and B.\\
\textit{Further work.} Ongoing effort focuses on the automation of all the chains from a P-Net tool to the B tools. We investigate the transformation into a B machine, of the XML outputs of the tools such as the PEP tool \footnote{{\small \texttt{sourceforge.net/projects/peptool}}}. The result is to be passed as the included machine. But, many experiments of various size are still needed for the scalability of our translation process. Meanwhile, user-friendly tools to facilitate the combination of the techniques are of a major interest. 
\vspace{-0.2cm}
\bibliographystyle{plain}

\input{pn2b.bbl}
\end{document}

%% file: ovv_eventB.tex
An \textit{abstract system} \cite{Abr96a,AbrialMussat98} describes a mathematical model of a system behaviour\footnote{A system behaviour is the set of its possible transitions from state to state beginning from an initial state.}. It is  mainly made of a state description (constants, variables and invariant) and several \textit{event} description. 
While \textit{abstract machines} are the basic structures of the earlier operation-driven approach of the B method, \textit{abstract systems} are the basic  structures of the so-called \textit{event-driven} B, and replace abstract machines. 
Abstract systems are comparable to Action Systems \cite{BaKu83}; they describe a nondeterministic evolution of a system through guarded actions. 
Dynamic constraints can be expressed within abstract systems to specify various liveness properties \cite{AbrialMussat98,ZB02-Cansell}. 
The state of an abstract system is described by variables and constants linked by an invariant. Variables and constants represent the data space of the system being formalized. Abstract systems may be refined like abstract machines \cite{ZB02-Cansell,AbrialCansellMery03}. 
\vspace{-0.4cm}
\subsubsection*{Data of an Abstract System} At a higher level an abstract system
models and contains the data of an entire model, be it distributed or not.
Abstract systems have been used to formalize the behaviour of vaios (including distributed) systems \cite{Abr96a,ButlerWalden96,ZB02-Cansell,AbrialCansellMery03}. Considering a global vision, the data that are formalized within the abstract system may correspond to all the elements of the distributed system.
\vspace{-0.4cm}
\subsubsection*{Events of an Abstract System} Within B, an event is considered as in the approach of Action Systems, \textit{i.e.} as the observation of a system transition. Events are  spontaneous and show the way a system evolves. An event has a \textit{guard} and an \textit{action}. It may occur or may be observed only when  its guard holds. 
The action of an event describes with generalized substitutions how the system state evolves when this event occurs.
Several events can have their guards held simultaneously; in this case, only one of them  occurs. The system makes internally a nondeterministic choice. If no guard is true the abstract system is blocking (deadlock).
\begin{figure}[htp]
\begin{center}
\begin{multicols}{2}

\noindent
\begin{boxedminipage}{5cm}
\begin{tabbing}
\hspace{0.4cm}\=\hspace{0.4cm}\=\hspace{0.4cm}\=\hspace{0.4cm}\kill
\>\texttt{name} $\defs$ \textsf{/* event name */} \\
\>\>\textsc{select}\\
\>\>\>     $P_{(gcv)}$ \\
\>\>\textsc{then} \\
\>\>\>     $GS_{(gcv)}$\\
\>\>\textsc{end}
\end{tabbing}
\end{boxedminipage}
\centerline{(SELECT Form)}
\noindent
\begin{boxedminipage}{5cm}
\begin{tabbing}
\hspace{0.4cm}\=\hspace{0.4cm}\=\hspace{0.4cm}\=\hspace{0.4cm}\kill
\>\texttt{name} $\defs$ \textsf{/* event name */} \\
\>\>\textsc{any} $bv$ \textsc{where}\\
\>\>\>     $P_{(bv, gcv)}$ \\
\>\>\textsc{then} \\
\>\>\>     $GS_{(bv,gcv)}$\\
\>\>\textsc{end}
\end{tabbing}
\end{boxedminipage}
\centerline{(ANY Form)}
\end{multicols}
\caption{General Forms of Events}
\label{figure:eventshape}
\end{center}
\end{figure}
An event has one of the general forms (Fig. \ref{figure:eventshape}) where $gcv$ denotes the global constants and variables of the abstract system containing the event; $bv$ denotes the bound variables (variables bound to \textsc{any}). $P_{(bv, gcv)}$ denotes a predicate $P$ expressed with the variables $bv$ and $gcv$; in the same way $GS_{(bv,gcv)}$ is a generalized substitution $S$ which models the event action using the variables $bv$ and $gcv$.
The \textsc{select} form is just a particular case of the  \textsc{any} form. The guard of an event with the \textsc{select} form is $P_{(gcv)}$.
The guard of an event with the \textsc{any} form is $\exists(bv).P_{(bv,gcv)}$.
\vspace{-0.4cm}
\subsubsection*{Semantics and Consistency.} An abstract system describes a mathematical model that simulates the  behaviour of a system. Its semantics arises from the invariant and is  guaranteed  by proof obligations (POs). The \textit{consistency} of the model is established by such proof obligations:  
\textit{i)} \textit{the initialisation should establish the invariant};  \textit{ii)} \textit{each event of the given abstract system should preserve the invariant of the model} (one must prove these POs). 
The proof obligation of an event with the \textsc{any} form  is: $$I_{(gcv)} \land P_{(bv,gcv)} \implies [GS_{(bv,gcv)}]I_{(gcv)}$$ where $I_{(gcv)}$ stands for the invariant of the abstract system.